# Special Relativity from Classical *Gedankenexperiments* involving Electromagnetic Forces: A Contribution to Relativity without Light

Sergio E S Duarte*

*Department of Physics, Federal Center of Techonological Education Celso Suckow da Fonseca, Rio de Janeiro - Brasil*

Nathan W Lima **

*Department of Physics, Federal Center of Techonological Education Celso Suckow da Fonseca, Rio de Janeiro - Brasil*

## Abstract

Simply by assuming the first postulate of Special Relativity and by exploring *Gedanken*experiments with electromagnetic forces, we suggest that there is a speed limit in the universe, which can be determined as a relation between vacuum permeability and permittivity. We also derive space contraction and time dilatation without referring to light. Finally, we argue that, based on previous works on the derivation of Lorentz's Transformation without light and from our results, it is possible to transform the epistemological status of the Einstein's second postulate from principle into derivation.

## I. Introduction

Special Theory of Relativity (STR) is one of the most significant revolutions in the history of science [1,2]. It has led to new conceptions of space and time, including the recognition of counter-intuitive phenomena, such as time dilatation and space contraction, which can be derived from Einstein's two postulates through Lorentz's Transformations.[3]

Einstein's first postulate states that "the same laws of electrodynamics and optics will be valid for all frames of reference for which the equations of mechanics hold good" and the second one states that "light is always propagated in empty space with a definite velocity *c* which is independent of the state of motion of the emitting body".[4] As long as science looks for universal descriptions of physical reality, the first postulate seems very reasonable and one finds no difficulty in accepting it *a priori*. On the other hand, assuming that there is a speed value which is independent of any other movement is not only counter-intuitive but also counter-inductive[5] as it transgresses Galileo's relativity. Michelson, for instance, kept challenging this postulate even when General Relativity was already stablished.[6–10] Furthermore, an important result of Quantum Physics, de Broglie's wave-particle duality, was articulated assuming that photons have mass, which implies that electromagnetic radiations with different frequency have different velocities.[11, 12]

Considering the difficulties of accepting Einstein's second postulate *a priori*, it is possible to find different works that propose Relativity without (the postulate of) light. By assuming space-time isotropy and homogeneity, one may achieve the general structure of the addition law for parallel velocities[13] and even the structure of Lorentz transformations in which a limit speed *c* shows up, but its value remains undetermined.[14–16]

In the present work, we intend to introduce a small contribution to the proposition of Relativity without light. We have three goals: first, exploring *Gedankenexperiments*, simply through force laws and the conservation hypothesis, we derive the existence of a speed limit *c* in nature, guaranteeing that different observers agree about the dynamic state of a system (equilibrium, repulsion or attraction). We show that *c* is determined by a relation between vacuum permittivity and permeability, finding its precise value. Our second goal is to derive space contraction and time dilatation without referring to light, from other *Gedankenexperiments*. Finally, our third goal is to indicate that the undetermined velocity in the Lorentz transformations obtained by "Relativity without Light" may be the limit speed we obtained. By doing so, we also argue that *c* is invariant under reference frame transformation.

To achieve these goals, we assume the following hypotheses: (i) Coulomb's Force, Lorentz's Force and Biot-Sarvat Law are valid for all inertial reference frames. (ii) The dynamic state of a system is the same in all inertial reference frames. Any two bodies are subject to three possible dynamic states: repulsion, attraction or equilibrium. If one observer verifies that two particles are repelling each other, for instance, it is reasonable to assume that a second observer in any other reference frame will also verify the same. Actually, this can be understood as a part of Einstein's Relativity Principle. And, (iii) electric charge is invariant under reference frame transformations. This is supposed to be valid as long as electric charge can be considered a measurement of particle quantity and its variation would imply the creation or annihilation of particles.

We understand that our proposal makes three contributions: (i) we obtain two counter-intuitive results (space contraction and time dilatation) and the existence of a speed limit (which is computed as a relation of two coupling constants) departing from force laws that are empirically well tested and accepted. We have not mentioned anything about the nature of light, which is (in contemporary physics) a much more recent and sophisticated topic than electric and magnetic forces. Thus, we provide empirical elements to articulate relativity, which historically has not been done[17] (ii) We change the epistemological status of the invariance of speed of light i from postulate to derivation. (iii) We introduce very basic *Gedankenexperiments* to report that a revision of space and time conceptions is already demanded by electromagnetic theory without the necessity of any further postulate.

## II. A speed limit in nature

We start by proposing a *Gedankenexperiment* in which two particles with positive charge are separated by a distance *d* along the x-axis. They have no relative velocity at t = 0. We shall examine the force applied in particle 2 in two different reference frames: S (proper reference frame), in which particles are at rest at t = 0, and S' (non-proper reference frame), in which particles are not at rest at t = 0. This is depictured in figure 1.

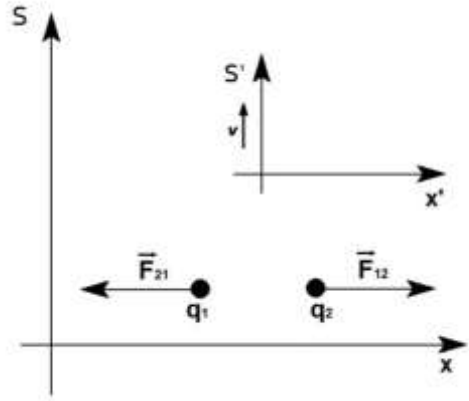

FIG. 1. Two particles with no relative velocity at t = 0. The situation is examined in two reference frames: S (proper reference frame) and S' (non-proper reference frame).

### a) Case I: Proper reference frame

In reference frame S, both particles are at rest at t = 0. In this case, the resultant force on particle two is written simply as a Coulomb force:

$$\vec{F_r} = \frac{1}{4\pi\epsilon_0} \frac{q_1 q_2}{d^2} \hat{x} \qquad (1)$$

$\vec{F_r}$ denotes the resultant force in S.

### b) Case II: Non-proper reference frame

In reference frame S', besides the Coulomb force, particle two is also object of a magnetic force, which can be computed using Lorentz's expression $\vec{F} = q\vec{v} \times \vec{B}$. In this case, q and $\vec{v}$ are the charge and velocity of particle 2, and $\vec{B}$ is the magnetic field created by the movement of particle 1. Thus, the resultant force may be written as:

$$\vec{F_r'} = \vec{F_e'} + \vec{F_m'} \qquad (2)$$

Where $\vec{F_r'}$, $\vec{F_e'}$ and $\vec{F_m'}$ are the resultant, electric and magnetic forces in S', respectively. Coulomb force is still the same:

$$\vec{F_e'} = \frac{1}{4\pi\epsilon_0} \frac{q_1 q_2}{d^2} \hat{x} \qquad (3)$$

And the magnetic force can be computed using Lorentz's Force and Biot-Sarvat's Law:

$$\vec{F_m'} = -q_1 v B \hat{x} = -q_1 v \frac{\mu}{4\pi} \frac{v q_2}{d^2} \hat{x} \qquad (4)$$

By defining a constant $c = \sqrt{\frac{1}{\mu_0 \epsilon_0}}$, we can rewrite $\mu_0 = \frac{1}{c^2 \epsilon_0}$ and substitute it in (4):

$$\vec{F_m'} = -\frac{1}{4\pi\epsilon_0} \frac{q_1 q_2}{d^2} \frac{v^2}{c^2} \hat{x} \qquad (5)$$

Thus, the resultant force in S' is

$$\vec{F_r'} = \frac{1}{4\pi\epsilon_0} \frac{q_1 q_2}{d^2} \left(1 - \frac{v^2}{c^2}\right) \hat{x} \qquad (6)$$

According to equation (6), an observer in reference frame S and S' do not agree on the intensity of the force that is applied on particle 2, except when the velocity of S' is zero in relation to S. When $v$ is different from zero, we identify three different cases, which are summarized in table 1.

TABLE 1. Dynamic State in difference reference frames

|  | Particles velocity | Dynamic State | Does it agree about the dynamic state with the proper-reference frame? |
|---|---|---|---|
| Proper reference frame | $v = 0$ | Repulsion. | - |
| Non-proper | $0 < v < c$ | Repulsion | Yes |
| Non-proper | $v = c$ | Equilibrium | No. |
| Non-proper | $v > c$ | Attraction. | No. |

Equation (6) and the acceptance of the Principle of Relativity imply that no reference system can travel with a speed equal or above $c$, otherwise observers will disagree about very simple and fundamental observations. For $v = c$, an observer in S will say that particles will move away from each other in the subsequent instant and an observer in S' will say that particles will not have any relative motion in the subsequent instant. For $v > c$, an observer in S will say that particles will move away from each other in the subsequent instant and an observer in S' will say that they will move into each other in the subsequent instant. This means that they will even disagree about what they are seeing without the necessity of any measurement.

It is important to notice, however, that, for $v < c$, both observers agree about the dynamic state of the system (so they do not have a dramatic disagreement about their observations of reality), but they still disagree about the module of the force that is applied in particle 2. In the next section, we examine a second *Gedankenexperiment* in order to deepen the comprehension of the implications of what electromagnetic theory brings when motion is taken into account.

### III. Space Contraction and Time Dilatation

In section II, we analyzed a particle that was not in equilibrium in the proper reference frame. In this section, we examine the other possible situation, a particle that is in equilibrium in the proper reference frame. Again, according to the Relativity Principle, if one observer verifies that a system is in equilibrium, any other observer should also verify it.

The *Gedankenexperiment* consists of a charged particle that is equidistant to two lines with charged density $\lambda_1$ and $\lambda_2$ (figure 2). In the proper reference frame S, the particle is at rest and in equilibrium. One of the lines has current to the right direction consistent with a charge velocity $v$. In a second reference frame S', which is moving with a velocity $v$ in relation to S, the particle is moving to the left with a velocity $v$, line one presents no current and line two presents current to the left.

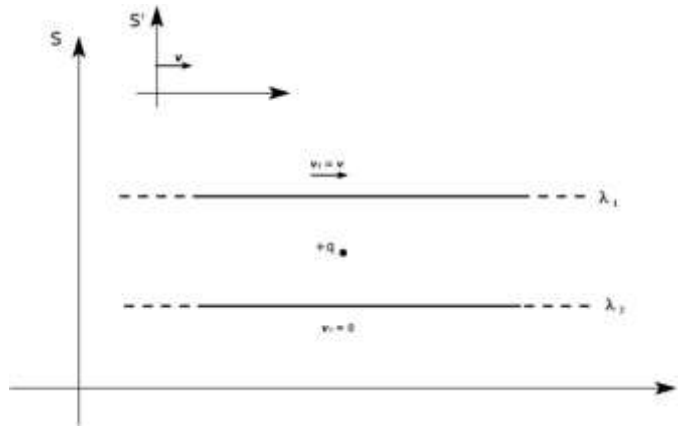

FIG 2. A charged particle is equidistant to two lines with charged density $\lambda_1$ and $\lambda_2$. In the proper reference frame S, the particle is at rest and in equilibrium. One of the lines carries current in the right direction, consistent with a charge velocity $v$. In a second reference frame S', with a velocity $v$ in relation to S, the particle is moving to the left with a velocity $v$, where line one carries no current and line two carries current to the left.

### a) Case I: Proper Reference Frame

In S the particle is at rest, not being susceptible to any magnetic force. Once the particle is in equilibrium, the electric force caused by line 1 must be equal to the electric force caused by line two. Using Coulomb's Law, we have that

$$|\vec{F_1}| = \frac{q\lambda_1}{2\pi\epsilon_0 r_1} \tag{7}$$

And

$$|\vec{F_2}| = \frac{q\lambda_2}{2\pi\epsilon_0 r_2} \tag{8}$$

Since $r_1 = r_2$, we conclude that the charge density of both strings is the same in the proper reference frame:

$$\lambda_1 = \lambda_2 = \lambda \tag{9}$$

### b) Case II: Non-Proper Reference Frame

In S', the particle is not at rest. Thus, it is sensible not only to the electric forces caused by both lines but also to the magnetic force caused by line two. Since we are assuming that the particle must be in equilibrium in all reference frames, we have that

$$|\vec{F'_{1el}}| - |\vec{F'_{1mag}}| = |\vec{F'_{2el}}| \tag{10}$$

The electric forces, which have the same value in the proper reference frame, must be different in S' in order to guarantee the equilibrium with the magnetic force. As for the particle is equidistant to the lines, the only parameter that may differ one line from the other is the charge density $\lambda$. Thus, we start to look for an expression for the charge density in reference frame S':

$$|\overrightarrow{F'_{1el}}| = \frac{q\lambda'_1}{2\pi\epsilon_0 r} \qquad (11)$$

And

$$|\overrightarrow{F'_{2el}}| = \frac{q\lambda'_2}{2\pi\epsilon_0 r} \qquad (12)$$

Using Biot-Savart Law and Lorentz's Force, we find that the magnetic force on the particle must be

$$|\overrightarrow{F'_{1mag}}| = \frac{\mu \lambda'_1 v^2 q}{2\pi r} \qquad (13)$$

Using $\mu_0 = \frac{1}{c^2 \epsilon_0}$, we obtain:

$$|\overrightarrow{F'_{1mag}}| = \frac{\lambda'_1 q v^2}{2\pi\epsilon_0 r} \qquad (14)$$

By substituting (11), (12) and (14) in (10), we find a relation between the new charge densities in S':

$$\lambda'_1 \left(1 - \frac{v^2}{c^2}\right) = \lambda'_2 \qquad (15)$$

Equation (15) leads us to a similar situation we faced in section II. The constant $c$ imposes a speed limit in nature. If v < c, although the value of charge density differs from one line to the other, the sign of the charge does not change. If v = c, the second line turns into an uncharged line. And, if v > c, the second line manifests an opposite charge to the one observed in the proper reference frame. In order to all observers agree about the nature of the charge density (positive, negative or neutral), no system may have $v$ equal or greater than c.

### c) Space Contraction

Until this point we have verified that observers with different velocities perceive different forces and different charge densities. We are now interested in finding the exact relation between the proper and non-proper quantities. In order to make the simplest possible hypothesis, let us suppose that the relation between the proper charge density and the non-proper charge density is linear, $\lambda_{prop} = \alpha \lambda_v$. Thus,

$$\lambda_1 = \alpha \lambda'_1 = \lambda \tag{16}$$

And

$$\lambda'_2 = \alpha \lambda_2 = \alpha \lambda \tag{17}$$

Substituting (16) and (17) in (15), we obtain

$$\frac{\lambda}{\alpha}\left(1 - \frac{v^2}{c^2}\right) = \alpha \lambda \tag{18}$$

This leads to:

$$\alpha = \sqrt{1 - v^2/c^2} \tag{19}$$

Thus, we can state that

$$\lambda_{prop} = \lambda_v \sqrt{1 - v^2/c^2} \tag{20}$$

While $\lambda_{prop}$ is the charge density in the proper reference frame, $\lambda_v$ is the charge density in a reference frame with velocity $v$ in relation to the proper reference frame. Considering that $\lambda = q/L$ and that q is invariant, the only property that may vary is the line length:

$$\frac{q}{L_{prop}} = \frac{q}{L}\sqrt{1 - v^2/c^2} \tag{21}$$

And, thus, we finally obtain:

$$L = L_{prop}\sqrt{1 - v^2/c^2} \tag{22}$$

This is the relation known as space contraction in STR. Thus, in order for a particle to be observed in equilibrium in all reference frames with velocity below *c*, it is necessary to assume that the lengths of the charged lines vary from one reference to the other. This result was obtained using only electromagnetic forces and the Relativity Principle. By assuming that length contraction is always observed in different reference frames, we may also obtain a traditional result of Special Relativity: time dilatation.

### d) Time Dilatation

From the last result we have obtained (space contraction), it is easy to show that time is also affected by movement without any further hypothesis. Suppose that two events happen in the same position at a reference frame S'. The time interval between these two events is $\Delta t'$. Assuming $t'_0 = 0 \rightarrow t' = \Delta t'$. As $t'$ is measuring time where the events are happening in the same position, we call it proper time. In a different reference frame S, in which S' is observed moving with a velocity v, the distance between the two events is measured (thus, in S we have the proper length). This is represented in Figure 3.

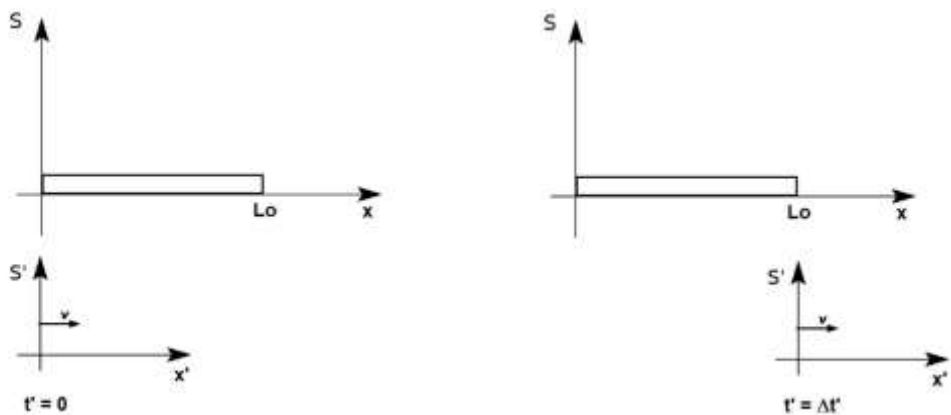

FIG 3. Two events happen in the same position at a reference frame S'. In reference

frame S, in which S' is observed moving with a velocity v, the distance between the two events is measured – corresponding to $L_0$, the proper leng.

Thus, we have that the proper-time t' and the time t are equal to:

$$t' = \frac{L'}{v} \quad \text{and} \quad t = \frac{L_0}{v} \tag{23}$$

And, we may relate time in both reference frames:

$$t = \frac{t' L_0}{L'} \tag{24}$$

As $L_0$ is the proper length, we have that $L' = L_0\sqrt{1 - v^2/c^2}$. By denoting t' as the $t_{prop}$ (proper time), we have:

$$t = \frac{t_{prop}}{\sqrt{1 - v^2/c^2}} \tag{25}$$

Which is the time dilatation expression of STR.

### IV. Lorentz's transformations, velocity addition rule and speed limit in nature

There is a vast literature on the derivation of Lorentz's Transformation without the postulate of the constancy of the speed of light [14]. Different approaches may be developed depending on the principles that are used, such as reciprocity principle, space-time homogeneity, spatial isotropy, relativistic transverse momentum conservation, group property, relativistic mass increase, causality postulate, uniqueness postulate [18].

In these derivations, a general form of Lorentz's transformation is obtained [15, 16] and, consequently, it is possible to derive the existence of a speed limit in the universe whose value is not intrinsically determined. Combining this idea with the results we obtained from classical electromagnetic forces, it is reasonable to assume that $c = \sqrt{\frac{1}{\mu_0 \epsilon_0}}$ is this speed limit. Furthermore, by using the relativistic expression for velocity addition (obtained from Lorentz's transformations), it is possible to find that whenever something has speed $c$ in a specific inertial reference frame, it will have speed $c$ in all inertial reference frames.

Considering that a mathematical equation can assume different epistemological status in a physical description, i.e., principle, empirical law, derivation and definition [19], we understand that this work conveys a last epistemological contribution, as it transforms Einstein's second postulate of relativity from principle into derivation.

## V.    Conclusions

According to Newton's Dynamics, force is proportional to the second derivative of position in relation to time. In this sense, two reference frames with constant relative velocity should perfectly agree in all their description about reality. As matter of fact, this is the physical principle that relies below Galileo's Principle of Relativity. The problem in this description, however, is already present in the Classical Theory of Electromagnetism. Lorentz's Force is proportional to velocity, implying that two observers with constant relative velocity disagree about the measurement of forces.

In this work, we departed from Classical Electromagnetism (without making any reference to the speed of light) to show that if a speed limit $c = \sqrt{\frac{1}{\mu_0 \epsilon_0}}$ is achieved or overcome, observers will disagree not only about the resultant force module but also about the dynamic state of systems (equilibrium, repulsion, attraction). We have shown that to overcome this velocity would also imply drastic alterations in the description of reality such as the inversion of the sign of the charge density of a system.

We have also demonstrated that, simply based on the Principle of Relativity and electromagnetic force laws, it is possible to derive space contraction and, consequently, time dilatation. This possibility reinforces Richard Feynmann's idea that electric and magnetic field are relativistic aspects of the same phenomena.[20] Finally, we have shown that, assuming that a general form of Lorentz's transformations can also be obtained without the concept of light, $c = \sqrt{\frac{1}{\mu_0 \epsilon_0}}$ may be the limit velocity predicted by it and its value is invariant under Lorentz's transformation, a fact that turns Einstein's second postulate into a derivation.

___________________________________________________________

\* seduart11@gmail.com

\*\* lima.nathan@gmail.com

14  S. Gao, "Relativity without Light: A Further Suggestion." 2017.

15  V. Berzi and V. Gorini, "Reciprocity Principle and the Lorentz Transformations," *J. Math. Phys.*, 10 (8), 1518–1524 1969.

16  P. B. Pal, "Nothing but relativity," *Eur. J. Phys.*, **24** (3), p. 315-319, 2003.

17  R. Martins, "Force measurement and force transformation in special relativity," *Am. J. Sci.*, vol. **50** (11), 1008–1011 1982.

18  J. H. Field, "A new kinematical derivation of the Lorentz transformation and the particle description of light," *Helv. Phys. Acta*, **70**, 542–564 1997.

19  R. Karam and O. Krey, "Quod erat demonstrandum: Understanding and Explaining Equations in Physics Teacher Education," *Sci. Educ.*, vol. **24**(5), 661–698 2015.

20  R. Feynman, R. Leighton, and M. Sands, *Feynman's Lectures on Physics*. Pasadena: California Institute of Technology, 2013.